%% file: main.tex
\documentclass[conference]{IEEEtran}
\usepackage{cite}
\usepackage{amsmath,amssymb,amsfonts}
\usepackage{algorithm}
\usepackage{algpseudocode}
\usepackage{graphicx}
\usepackage{textcomp}
\usepackage{adjustbox}
\usepackage[utf8]{inputenc} 
\usepackage[T1]{fontenc}    
\usepackage{hyperref}       
\usepackage{url}            
\usepackage{booktabs}       
\usepackage{nicefrac}       
\usepackage{microtype}      
\usepackage[acronym]{glossaries}
\usepackage{tabularx}
\usepackage{cleveref}
\usepackage{glossaries}
\Crefformat{figure}{#2Fig.~#1#3}
\Crefmultiformat{figure}{Figs.~#2#1#3}{ and~#2#1#3}{, #2#1#3}{ and~#2#1#3}

\title{Defending against Stegomalware in Deep Neural Networks with Permutation Symmetry}
\author{
\IEEEauthorblockN{Birk Torpmann-Hagen}
\IEEEauthorblockA{\textit{UiT the Arctic University of Norway}\\
Tromsø, Norway \\
birk.s.torpmann-hagen@uit.no}
\and
\IEEEauthorblockN{Michael A. Riegler}
\IEEEauthorblockA{\textit{SimulaMet} \\
Oslo, Norway \\
michael@simula.no}
\and
\IEEEauthorblockN{Pål Halvorsen}
\IEEEauthorblockA{
\textit{SimulaMet}\\
Oslo, Norway \\
paalh@simula.no}
\and
\IEEEauthorblockN{Dag Johansen}
\IEEEauthorblockA{\textit{UiT the Arctic University of Norway} \\
Tromsø, Norway \\
dag.johansen@uit.no}
}

\input{glossaries.tex}
\begin{document}
\maketitle
\begin{abstract}
Deep neural networks are being utilized in a growing number of applications, both in production systems and for personal use. Network checkpoints are as a consequence often shared and distributed on various platforms to ease the development process. This work considers the threat of neural network stegomalware, where malware is embedded in neural network checkpoints at a negligible cost to network accuracy. This constitutes a significant security concern, but is nevertheless largely neglected by the deep learning practitioners and security specialists alike. We propose the first effective countermeasure to these attacks. In particular, we show that state-of-the-art neural network stegomalware can be efficiently and effectively neutralized through shuffling the column order of the weight- and bias-matrices, or equivalently the channel-order of convolutional layers. We show that this effectively corrupts payloads that have been embedded by state-of-the-art methods in neural network steganography at no cost to network accuracy, outperforming competing methods by a significant margin. We then discuss possible means by which to bypass this defense, additional defense methods, and advocate for continued research into the security of machine learning systems. 
\end{abstract}

\section{Introduction}
Stegomalware is a form of malware which utilizes steganographic techniques in order to impede detection~\cite{stegomalware_trends}. Typically, this entails concealing a malicious payload, for instance executable code or sensitive data for exfiltration, in otherwise inconspicuous files, such as image- or document-files~\cite{stegomalware_survey}. This is considered a significant emerging threat to cybersecurity in large part due to the difficulty involved in detecting and mitigating instances of stegomalware. Though there is emerging work towards addressing these concerns through developing detection methods based on, for instance, statistical heuristics, machine-learning, or other forms of static analysis~\cite{stegomalware_detection}, equivalent advancements are being made towards novel steganographic methods that evade detection.

Notably, recent work has shown that it is possible to embed stegomalware in deep neural networks at negligible cost to the network’s performance~\cite{maleficnet, stegonet,evilnet}. This poses a novel and significant security concern compared to other types of files in that the distribution of neural network weights cannot as easily be characterized compared to more conventional stegomalware media, and that neural networks permit the embedding of significantly larger payloads due to the sheer data volume required by neural network checkpoints. This means that attacks that utilize these methods are not only more difficult to detect, but can be more sophisticated and effective due to the increased capacity. This, coupled with the relative ease of distributing neural network checkpoints on platforms such as Pytorch Hub, HuggingFace or other hosting services, means that neural network stegomalware can conceivably become a widespread and serious threat with the continued proliferation of machine learning technologies. 

Moreover, while conventional stegomalware typically requires a second piece of compromised software to be extracted and executed, it has been shown that neural network stegomalware can be extracted and executed through the exploitation of existing vulnerabilities in deep learning implementation frameworks, notably insecure deserialization~\cite{maleficnet, vulnerabilities}. As deep learning systems often comprise very large code-bases with numerous dependencies~\cite{industrial}, other such vulnerabilities may be plentiful. As neural networks are often running on enterprise servers and advanced computing infrastructure, such an attack could incur significant costs, with the attackers potentially taking control of the servers, accessing sensitive data, or otherwise disrupting regular operations. 

To illustrate the potential severity of this attack-vector, suppose that, as part of military research, a nation's military implements autonomous navigation functionality for a weaponized drone. In the interest of robustness, they utilize a neural network pre-trained on ImageNet as the backbone of their model. This is a relatively standard practice in production systems, as training from scratch on such large datasets requires extensive computational resources. Suppose they naively download a pre-trained model from HuggingFace, TorchHub, or similar platforms, and that an adversary has embedded a malicious payload in the parameters of this model. As this payload is, given current state-of-the-art methods~\cite{maleficnet}, robust to fine-tuning and model-pruning, it is not at all unlikely that this payload makes it onto a deployed drone. Vulnerabilities in deep learning implementation libraries may then be exploited, extracting and executing the payload in accordance to a programmed trigger condition. Given a sufficiently sophisticated adversary, this could entail the complete take-over of the system given sufficient knowledge of the drone's interfaces. A more rudimentary attack could be to overload or otherwise disrupt the system, the consequences of which would regardless be severe. 

At the time of writing there are no publicly available methods that successfully detect stegomalware in deep learning models, nor any effective, efficient, or reliable means of neutralizing the threat through payload corruption. Though it is possible to corrupt naive attacks by utilizing network pruning or continuing to train the model after downloading it, state-of-the-art implementations are made robust to these defenses through utilizing error-correcting codes~\cite{maleficnet}. Certain model-compression methods\cite{model_compression} have shown promise in this regard, but this typically requires computing time and machine learning expertise in excess of what can be reasonably expected in the majority of contexts. As a result, these methods are not likely to be utilized in all but the most security-conscious development teams, as the benefits with respect to workflow and resource costs gained from downloading a pretrained model would be immediately offset by the need for implementing and running compression algorithms. 

To address this, we propose a highly effective defense against the current state-of-the-art in neural network stegomalware: \textbf{weight permutation}. We exploit the permutation symmetry of common neural network layers to shuffle the order of the entries in the weights and biases, effectively and reliably corrupting payloads embedded using state-of-the-art methods at no cost to network performance.

We summarize our contributions as follows:
\begin{itemize}
    \item We introduce \textit{weight permutation}, an effective and robust defense against neural network stegomalware.
    \item We show mathematically that this operation maintains functional equality in the network.
    \item We compare weight permutation to competing methods, namely model retraining and pruning, demonstrating empirically the efficacy of our method towards mitigating the threat of neural network stegomalware.
    \item We outline potential avenues of research for the development of permutation-invariant steganographic attacks.
\end{itemize}

We organize this work as follows: in \Cref{ch:related_work}, we review related work in the field. In \Cref{ch:method}, we introduce weight permutation, which we validate empirically in \Cref{ch:experiments}. We describe our methodology in \Cref{ch:methodology} Finally, we discuss the implications of our work and avenues of further research in \Cref{ch:discussion}.

\section{Related Work}\label{ch:related_work}
In this section, we outline related work on stegomalware, defenses, and mitigation strategies thereof, and recent work on neural-network stegomalware. 
\subsection{Stegomalware}
Malware refers to malicious software intentionally designed to cause damage to a(nother) computer system. With the proliferation of Internet-connected devices, malware has become a pervasive threat to modern computing systems, and can lead to data breaches, financial losses, and damage to critical infrastructure. Modern approaches to cybersecurity are in large part successful at mitigating the risk of malware, notably through sophisticated anti-malware solutions like heuristic analysis, sandboxing, and signature-based detection~\cite{malware_detection}. However, these methods often rely on detecting known patterns or behaviors, making them vulnerable to novel or obfuscated malware. An increasing portion of emerging research and development has thus been concerned with the implementation of novel obfuscation techniques as well as methods by which obfuscated malware can be detected. Notable examples of this include polymorphic and metamorphic malware~\cite{polymorph_malware}, where the malware modifies itself in order to evade signature-based matching, and stegomalware~\cite{stegomalware_survey}, where the malware is embedded in otherwise inconspicuous data-structures such as images, document files, audio, etc. 

The efficacy of steganography for obfuscation is exemplified by many recent instances of successful stegomalware proliferation. Stegoloader~\cite{stegoloader}, for instance, is a Trojan that works by downloading a PNG file containing malicious code onto the user's computer using a key-generation executable downloaded by the user in an attempt to pirate software. The payload embedded in the PNG file is then extracted and executed by the key-generation executable, infecting the user with ransomware. SyncCrypt~\cite{synccrypt} works in a similar manner, but is distributed through emails posing as court orders. Once a user downloads the infected file, a PNG file containing an embedded zip file with malware is downloaded, extracted, and executed. Concerningly, this attack was only successfully detected by one type of antivirus software in a test consisting of 58 different vendors. More sophisticated attacks are also greatly aided by steganographic techniques, therein Daserf~\cite{daserf}, a backdoor designed for cyber-espionage attacks against Japanese corporations which used steganographic techniques to conceal configuration files.

The difficulty in detecting and mitigating stegomalware is a growing concern, with a growing number of malware instances utilizing steganographic techniques~\cite{stegomalware_trends, stegomalware_article}. Choudhary~\cite{stegomalware_detection} and Chaganti et al.~\cite{stegomalware_survey} survey the landscape of known steganographic techniques, successful instances of stegomalware, and existing methods of stegomalware detection. They show that there exists a wide variety of different steganographic techniques, though the most common approaches involve modifying bits that do not significantly affect the perceptual qualities of the media for which they are intended. This includes modulating the least significant bits, embedding data in high-frequency areas of the Fourier domain and similar transforms. Spread spectrum techniques are also often utilized, where the data is spread evenly across the entire frequency spectrum with low amplitude. They then outline the various approaches to detecting instances of stegomalware - referred to as \textit{steganalysis}, which Chaganti et al. ~\cite{stegomalware_survey} categorize as either rich-model approaches or machine-learning approaches, where the former utilizes engineered features and heuristics and the latter trains machine learning models to discriminate between natural data and data which have been embedded with a payload using steganographic techniques. Though many of these methods are capable of successfully detecting stegomalware, they do not readily generalize across stenographic methods and mediums, nor do they report particularly high accuracies. Generally, steganalysis is dependent on the assumption that the medium in question exhibits some characteristic properties that can be modified if steganographic methods are applied. This not only depends on the medium, but the payload can be made to appear as random noise if it is encrypted prior to embedding. 

\subsection{Stegomalware in Neural Networks}
Though stegomalware currently already poses a significant threat to computer security, the parallel rise of deep neural networks is further compounding the risks. In addition to being useful with regards to the development of sophisticated steganographic embedding techniques~\cite{gan_stego}, it has been shown that neural networks can be particularly effective cover media for malicious payloads~\cite{maleficnet, evilnet, stegonet}. 

Reference~\cite{stegonet} was the first to investigate the feasibility of utilizing neural networks as cover media. They utilize least significant bit substitution to embed a malicious payload, which they claim can be extracted and executed using an insecure deserialization exploit in combination with a logit-based trigger condition. They show that this approach remains undetected by several steganalysis tools. This does, however, impede the network performance to a certain extent. Reference~\cite{evilnet} thus expands on this framework and outlines a more sophisticated embedding method that preserves the initial models' performance. Reference~\cite{maleficnet} notes the sensitivity of the aforementioned approaches to common backdoor-removal techniques such as network pruning and retraining~\cite{pruning_defence} and implements a more robust embedding scheme using error-correcting codes and spread-spectrum coding. In addition to increasing the robustness of the payload to these removal techniques, they also show that this incurs negligible cost to network performance and remains undetected by state-of-the-art steganalysis methods. 

Maleficnet~\cite{maleficnet} in particular is a highly effective, stealthy and robust form of neural network stegomalware, which currently lacks any methods by which it can be detected or mitigated. This constitutes a significant risk, in particular when coupled with the vulnerability of insecure deserialization inherent to most machine learning libraries. Realistically, all that is needed in order for an attacker to infect a target with arbitrary malware is to have them load the right checkpoint. In order to ensure the safety of neural network applications, it is necessary to develop methods that sufficiently mitigate these types of attacks.  

\section{Permutation Defense}\label{ch:method}
In this section, we outline the theory and implementation of weight permutation, the first method which effectively mitigates the risk of state-of-the-art neural network stegomalware. This defense leverages the permutation symmetry intrinsic to neural networks to partially shuffle the network's weights in memory, effectively corrupting any steganographic payload that an adversary may have embedded. We first outline how neural network stegomalware works and how the payload is distributed in the network weights. We then discuss the permutation symmetry of neural networks and how this symmetry can be leveraged towards corrupting stegomalware payloads by permitting the partial shuffling of weight matrices, requiring only the addition of re-permuting forward-hooks to be functionally identical to the original infected model. 

Asides from refraining from downloading pre-trained checkpoints altogether, there are in general two possible approaches to mitigating the threat of neural network stegomalware: either one has to detect that a given checkpoint contains a malicious payload prior to loading trough steganalysis, or the payload has to be removed before it has a chance of being executed. There are currently no successful steganalysis methods for neural network stegomalware, and although progress is being made towards general-purpose methods, these are generally incapable of reliably detecting instances of stegomalware in neural networks~\cite{evilnet}. Moreover, these methods generally do not tend to exhibit high degrees of certainty, but are rather treated as classifiers and evaluated accordingly. We thus contend that removal methods stand the greatest chance of mitigating the threat. Though network pruning and retraining has been shown to successfully remove early instances of stegomalware, state-of-the-art methods such as Maleficnet~\cite{maleficnet} are robust to these removal methods. Nevertheless, we contend that removal is the most viable means of mitigating these sorts of threats in a robust and effective manner. 

\begin{figure*}[htb]
    \centering
    \includegraphics[width=0.8\linewidth]{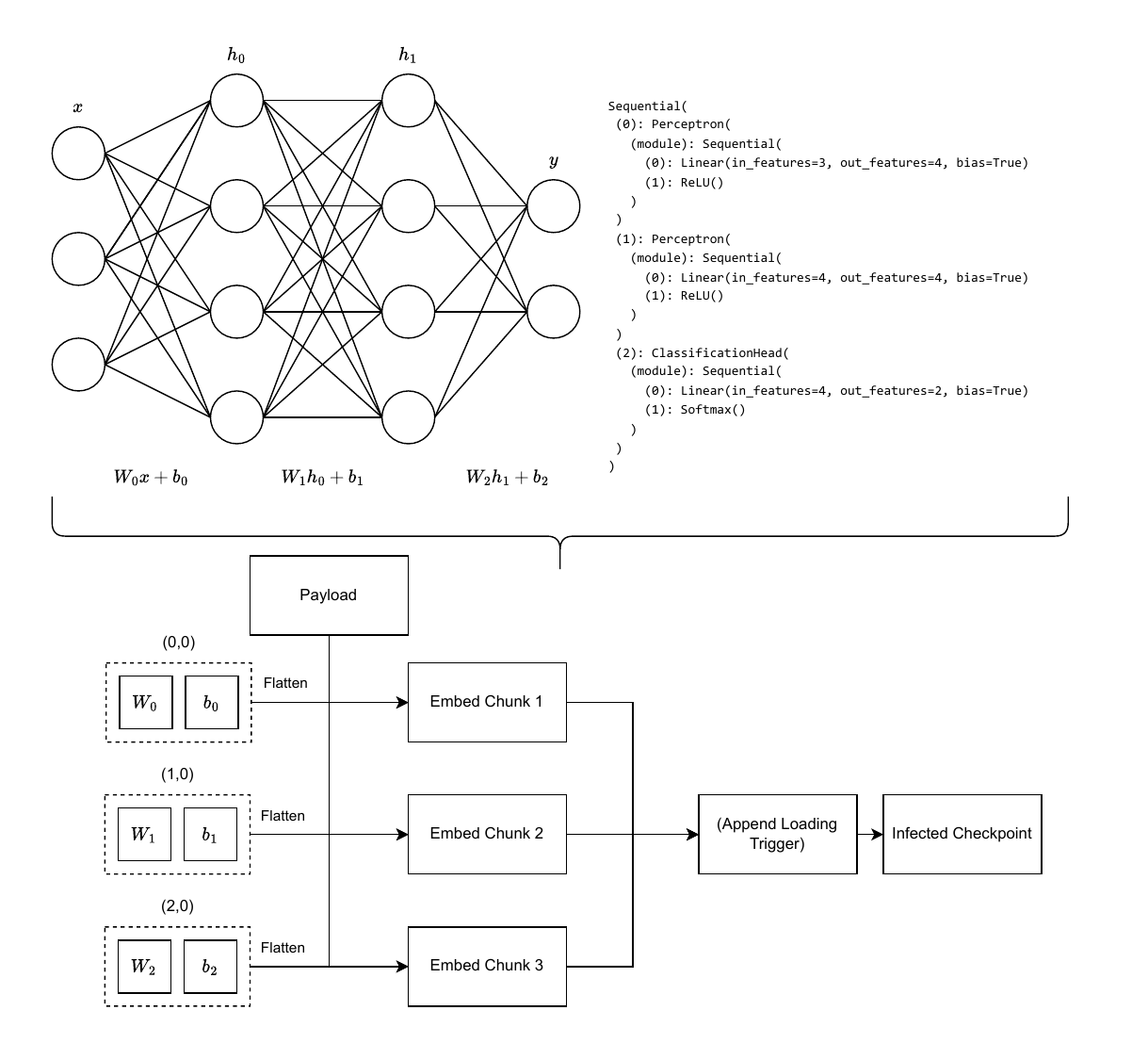}
    \caption{Visualisation of how neural network steganography works. Suitable weight- and/or bias-tensors for embedding are extracted from a neural network for each layer. A corresponding chunk of the payload is then injected into the flattened weight- and bias-tensors. One can then optionally append trigger code using deserialization vulnerabilities inherent to many machine-learning frameworks and distribute the infected checkpoint.}
    \label{fig:nn_memory}
\end{figure*}

To this end, it is first necessary to understand broadly how the payload is injected, extracted, and stored in neural networks. Current approaches to neural network steganography leverage the logical structure of neural networks to distribute the payload. In most machine-learning libraries, neural networks are stored in a large ordered dictionary, namely the state dictionary. Iterating over this dictionary yields keys that denote the layer weights and biases and values corresponding to the weight- and bias- tensors themselves, in sequential order from the input layer to the output layer.  Injecting the payload thus typically involves:
\begin{itemize}
    \item Splitting the payload into chunks,
    \item iterating over the state dictionary, and
    \item for each layer, embedding the corresponding payload chunk within the weight- and bias-tensors according to the embedding scheme, whether it is bit-substitution, spread-spectrum coding, or other methods.
\end{itemize}
The process is illustrated for a simple multi-layer Perceptron in \Cref{fig:nn_memory}.

Payload extraction is performed in much the same way: iterating over the layers, extracting and flattening the weight- and -bias tensors, and decoding the payload. Effective removal requires modifying the network such that the payload is irrecoverable. Thus, either the weight- and bias-tensors must be modified sufficiently, or there must be significant changes to the network architecture such that extraction becomes impossible. In either case, it is necessary to maintain the original performance of the network, as the process has to be applied indiscriminately to all checkpoints in order to constitute an effective security measure. 

Modifying the network architecture and thus the contents of the state-dictionary is fairly common practice, often performed for instance in order to adapt a pre-trained network to a new task or domain. This is, however, typically limited to the last few layers of the network, which can be altogether ignored in the embedding process if required. While more extensive modifications such as adding or removing intermediate layers or otherwise heavily modifying the architecture may likewise corrupt the payload, we contend that this is not a sufficiently robust defense. First and foremost, these sorts of changes are likely to adversely affect the network performance and thus counteract the benefits of downloading pre-trained checkpoints in the first place. Secondly, as the payload is distributed layer-wise, it is a relatively trivial matter to modify the extraction routine to only consider payload chunks which contain a predetermined hash, eliminating any effects from architectural modification.  

Consequently, it is necessary to modify the weights themselves in such a manner that the payload is corrupted. Network re-training and pruning, for instance, is an effective defense against earlier versions of stegomalware because it modifies the weights to an extent sufficient to render the payload irrecoverable. With error-correcting codes, however, these methods are rendered ineffective since the resulting changes are not particularly significant~\cite{maleficnet}. 

It is, however, possible to make significant changes to the weight- and bias-tensors without altering the functional nature of the neural network at all. This can be achieved by exploiting the fact that most neural network layers are permutation-symmetric, and thus that it is possible to freely permute the weight- and bias-tensors so long as the output is re-permuted accordingly. This property forms the basis of our method. By randomly permuting the network weights, the network is altered to an extent such that the payload is irrecoverable in spite of any error-correction methods. 

We can prove the permutation symmetry of neural networks as follows:

\textbf{Lemma 1}: linear layers are symmetric with respect to column order. 
\textbf{Proof}:
For each layer given by a weight-matrix \(\textbf{W}_i\), bias \(b_i\), and element-wise activation function \(\sigma(\cdot)\):
\begin{align}
    y_i &= \mathbf{W}_i y_{i-1}+\mathbf{b}\\
    \mathbf{P}^T\mathbf{P} y_i &= \mathbf{P^T}\mathbf{P}(\mathbf{W}_i y_{i-1}+\mathbf{b})\\
    y_i &= \mathbf{P^T}(\mathbf{P}\mathbf{W}_i y_{i-1}+\mathbf{P}\mathbf{b})
\end{align}
By induction, this also generalizes to layers that are composed of several matrix-multiplications and additions, therein recurrent units and attention layers, so long as each weight- and bias-tensor undergo the same permutation. 

\textbf{Lemma 2}: Convolutional layers are symmetric with respect to channel order.Let:
\begin{itemize}
    \item $\mathbf{X} \in \mathbb{R}^{C_{\text{in}} \times H \times W}$ be the input tensor, where $C_{\text{in}}$ is the number of input channels, and $H \times W$ is the spatial dimension.
    \item $\mathbf{W} \in \mathbb{R}^{C_{\text{out}} \times C_{\text{in}} \times k_h \times k_w}$ be the convolutional weight tensor, where $C_{\text{out}}$ is the number of output channels, and $k_h \times k_w$ are the kernel height and width.
    \item $\mathbf{y} \in \mathbb{R}^{C_{\text{out}} \times H' \times W'}$ be the output tensor after applying the convolution.
\end{itemize}
The convolution operation can be expressed as:
\[
\mathbf{y}_c(i,j) = \sum_{d=1}^{C_{\text{in}}} \sum_{u=1}^{k_h} \sum_{v=1}^{k_w} \mathbf{W}_{c,d}(u,v) X_d(i+u, j+v)
\]
for each output channel $c = 1, \dots, C_{\text{out}}$, where $Y_c$ is the output in the $c$-th channel, and $X_d$ is the input from the $d$-th channel.
Let $\pi$ be a permutation of the output channels. If we permute the weight tensor such that:

\[
\mathbf{W}'_{c,d}(u,v) = \mathbf{W}_{\pi(c),d}(u,v)
\]

the new output $Y'$ using the permuted weight tensor becomes:

\[
Y'_c(i,j) = \sum_{d=1}^{C_{\text{in}}} \sum_{u=1}^{k_h} \sum_{v=1}^{k_w} \mathbf{W}_{\pi(c),d}(u,v) \mathbf{X}_d(i+u, j+v)
\]
Now, let us re-permute the output according to the inverse of the permutation $\pi^{-1}$. Define the final output as:

\[
\tilde{Y}_c(i,j) = Y'_{\pi^{-1}(c)}(i,j)
\]

Substituting $Y'$ into the above equation, we get:

\[
\tilde{Y}_c(i,j) = \sum_{d=1}^{C_{\text{in}}} \sum_{u=1}^{k_h} \sum_{v=1}^{k_w} \mathbf{W}_{\pi^{-1}(\pi(c)),D}(u,v) \mathbf{X}_d(i+u, j+v)
\]
Since the permutation $\pi$ and its inverse cancel out in the indices, this simplifies to:
\[
\tilde{Y}_c(i,j) = \sum_{d=1}^{C_{\text{in}}} \sum_{u=1}^{k_h} \sum_{v=1}^{k_w} \mathbf{W}_{c,d}(u,v) \mathbf{X}_d(i+u, j+v)
\]

\begin{figure*}[htb]
    \centering
    \includegraphics[width=\linewidth]{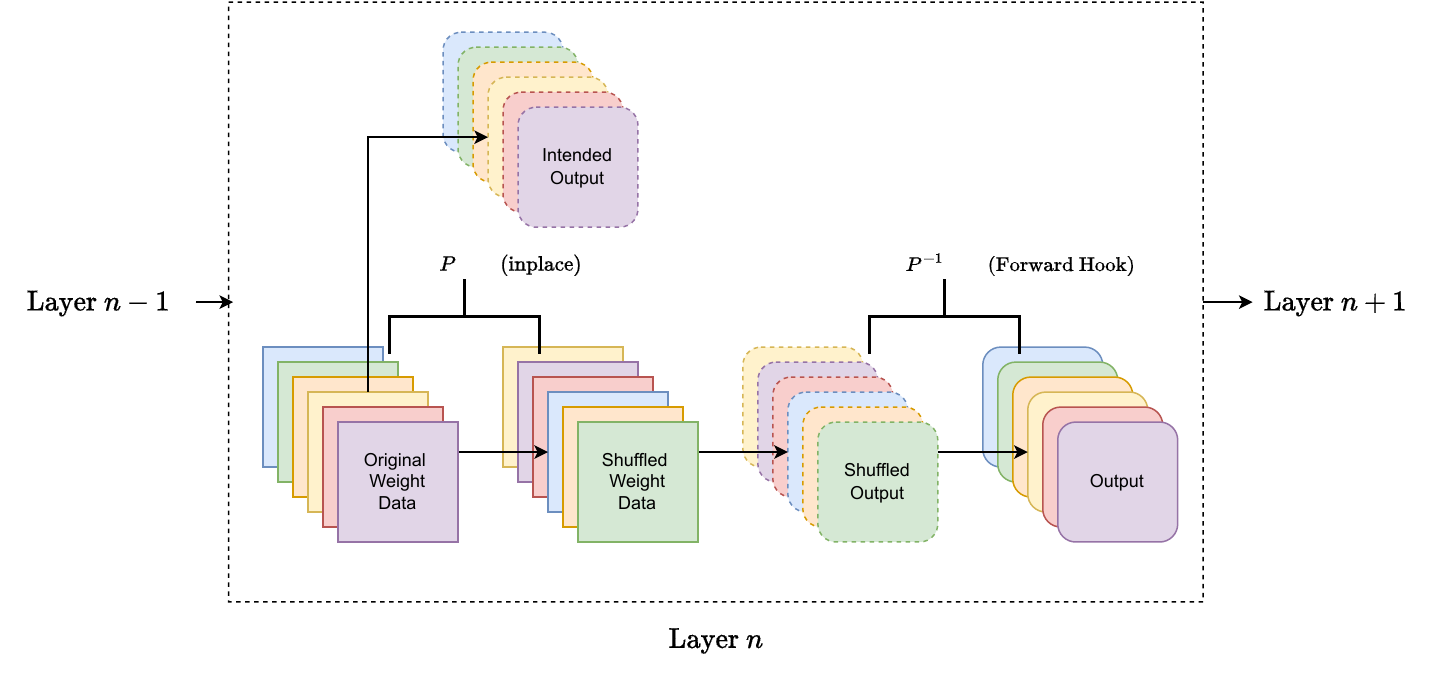}
    \caption{For each layer in the model with permutation-symmetric weights, we shuffle the weights along a symmetric axis and register a forward hook which repermutes the output into the intended order for the next layer.}
    \label{fig:channel-shuffling}
\end{figure*}

We have now shown that it is possible to freely permute the weights of both linear layers (and in general, layers consisting of matrix-multiplications) and convolutional layers while preserving their functional nature of so long as the output is re-permuted accordingly. This can be implemented as a forward hook at each layer output, as illustrated in \Cref{fig:channel-shuffling}. By induction, it is possible to apply this operation to each layer in the network, permuting the weight- and bias-data of the network as a whole while maintaining functional equality. Other architectural components, such as residual connections, batch normalizations, etc., do not affect this process, and custom layers, such as attention-layers and recurrent layers, can be implemented to support permutations so long as they are composed of matrix-multiplications or convolutions.

weight permutation works since current approaches to neural network stegomalware operate on flattened tensor representation of each layer's weights. Weight permutation shuffles the entries of these flattened tensors along the column/channel index, and thus makes it so that each entry of the shuffled tensor differs from the original tensor. While state-of-the-art methods do implement error-correction methods, notably low density parity checks in MaleficNet~\cite{maleficnet}, this can only account for a certain degree of noise, such as that caused by re-training and pruning, rather than the kind of high-impact structural changes caused by weight permutation. Indeed, weight permutation corresponds to a codeword error rate of 100\%. For the payload to be successfully extracted in this case, it would be necessary to re-order the tensors prior to extraction. Without knowledge of what permutation has occurred, it is thus impossible for the current state-of-the-art approaches to extract the payload. 

Weight permutation can thus be applied to any neural network where matrix multiplications are performed. This includes feed-forward networks, networks with self-attention such as large language models~\cite{transformers}, and recursive neural networks. Weight permutation can conceivably be implemented as default behaviour as part of the loading procedure in machine learning libraries in order to increase security, though as we will show in \Cref{ch:experiments} the required forward-hooks would increase execution times in this case.

We summarize our algorithm for weight permutation in \Cref{alg:wp}.
\input{figures/algorithm.tex}
\section{Methodology}\label{ch:methodology}
In this section, we outline the experimental setup we used to validate the efficacy of our methods, including our choice of baselines, test-bed, and metrics. We provide all the data used in this paper as well as comprehensive instructions on how to replicate our experiments on our GitHub page\footnote{\url{https://github.com/BirkTorpmannHagen/PermutationDefence}}.

We assess our methods against the current state-of-the-art in neural network steganography, MaleficNet~\cite{maleficnet}. There are currently no competing methods of neural network stegomalware mitigation beyond what the MaleficNet authors have discussed, namely network pruning, retraining, or student-teacher learning. These strategies were only discussed insofar that they were insufficient, however, with the methods they introduce being shown to be robust to both pruning and retraining and arguing that more sophisticated methods such as knowledge-distillation, though effective, would be an infeasible defense given a realistic threat scenario due to the computational resources required generally being approximately comparable to training from scratch~\cite{maleficnet}. Nevertheless, we utilize network pruning and retraining as our baselines. We perform 5\%, 10\%, 25\%, 50\%, 75\%, 90\%, and 95\% global unstructured pruning and retrain for 1, 5, 25, and 50 epochs. 

Since both weight permutation and stegomalware can be applied to any arbitrary neural network, as long as they are composed of matrix multiplications or convolutions, we focus on a single architecture, ResNets~\cite{resnet}, for simplicity of presentation. We emphasize that neither weight permutation nor neural network stegomalware depends on the architecture itself. These techniques only require access to the weight and bias tensors of the constituent layers. While different architectures may influence how payloads respond to pruning or retraining, exploring these interactions across specific architectural components lies beyond the scope of this paper.

An important aspect of any mitigation strategy is that they must not cause significant degradation to the network's accuracy. We thus also quantify the degree to which network accuracy is maintained after the mitigation methods have been applied. To this end, we train the ResNets on CIFAR10 and compute an accuracy quotient which represents the proportion of the original accuracy which is maintained after modifying the network. In order to get a complete picture of the degree to which the removal methods affect network performance and payload integrity, we test our methods for a range of neural network parameter counts. In particular, we consider every packaged size of ResNet in Pytorch: ResNet18, ResNet34, ResNet50, ResNet101, and ResNet152. We assess our methods across payload sizes of 100, 1024, and 10240 bytes. 

Finally, as a practically applicable mitigation strategy must also be computationally efficient, we include execution times for weight permutation. This includes the execution time necessary to reorder the weights as well as the overhead induced by the need for adding forward hooks in order to correctly re-permute each layer's outputs. We also include an analysis of how many weights require permutation in order to sufficiently corrupt the payload.

\section{Results}\label{ch:experiments}
In this section, we outline the results of our experiments. Overall, we observe that weight permutation effectively and efficiently corrupts the payload at no cost to network accuracy, albeit with some computational overhead. Though retraining and pruning in some cases also manages to corrupt the payload, the former only achieves this in one instance at 50 epochs, and the latter only after a degree of pruning that induces a significant loss of accuracy. We further show that the computational overhead caused by the permuting forward hooks can be moderately reduced, however, by selectively permuting only a set proportion of the weights, though this is at some cost to security. 

Our first experiments consist of an analysis of the neural networks' performance and payload integrity using different defenses. We compare our methods to network pruning and retraining. the results of which are shown in \Cref{tab:integrity}. Weight permutation effectively corrupts the payload across all models and payload lengths with no cost to accuracy. Global unstructured pruning does not adversely affect neither performance nor integrity to a significant extent until a pruning rate of 75\%. From this point the payload gets corrupted, albeit at a significant cost to network accuracy. Pruning is thus an infeasible defense, as sufficiently corrupting the payload requires pruning the model to such an extent that it no longer performs as required. Network retraining does not impede performance, but also fails to corrupt the payload in all cases except for training ResNet50 for 50 epochs. 
\input{figures/integrity}. 

While weight permutation clearly outperforms the competing methods, it incurs a degree of computational overhead. This includes an overhead with respect to loading the model and executing the model due to the additional forward hooks. A breakdown of the execution time of weight permutation for a range of ResNet sizes is shown in \Cref{tab:performance}. The loading overhead is unlikely to adversely affect production systems, as the loading process only occurs once. Nevertheless, our results indicate that this overhead is negligible, at most adding 356ms. The execution overhead is more concerning, however, ranging between increases of 26\% and 48\%. In systems where the execution time is of particular salience, such as in real-time systems, this overhead may be too large to warrant the added degree of safety from neural network stegomalware. 
\input{figures/performance}

\begin{figure}
    \centering
    \includegraphics[width=\linewidth]{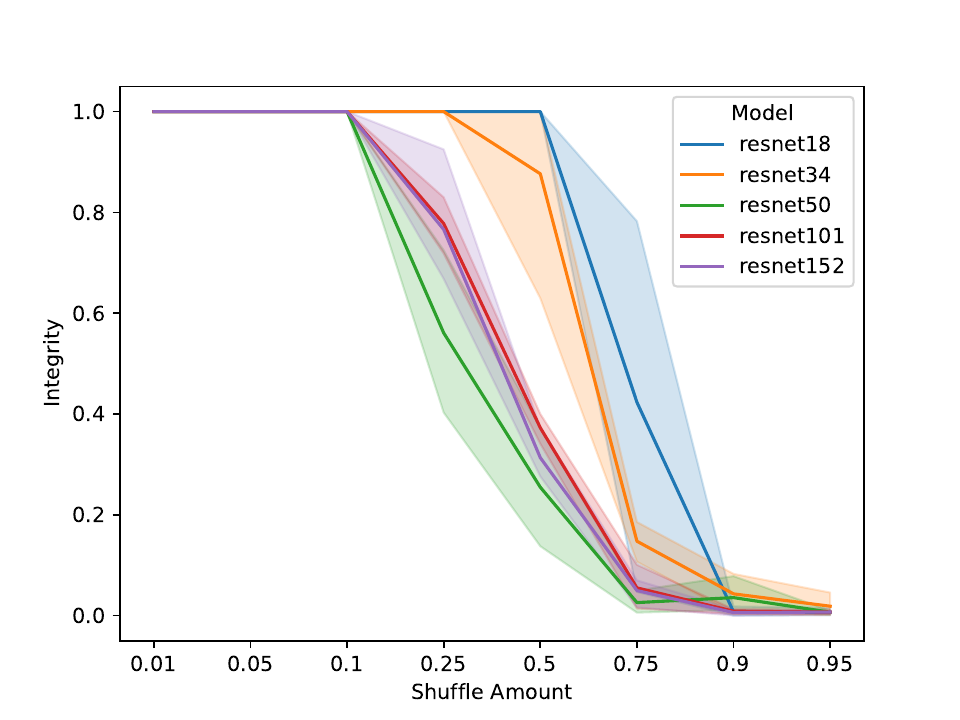}
    \caption{Plot showing the relationship between the proportion of shuffled weights vs the resulting payload integrity across model sizes.}
    \label{fig:integrity_sweep}
\end{figure}

This overhead is caused entirely by the need for forward hooks that re-permute the output of each layer. This overhead can thus be reduced by shuffling the order of one layer's weights and permuting subsequent layers accordingly, thus only requiring one single final forward hook. However, this reduces the safety since a savvy attacker can conceivably implement an extraction routine which checks several permutations. Another approach is to selectively permute only certain layers in proportion to overhead tolerances. To assess whether this still sufficiently corrupts the payload, we analyzed the payload integrity after permuting a range of different proportions of the network. The results are shown in \Cref{fig:integrity_sweep}. For the larger models, it is sufficient to permute around 25\% of the weights. Smaller models appear to require more permutation, typically around 75\%. It is, however, worth noting that reconstruction may still be possible with a partially corrupted payload, and hence, the proportion of weights to be permuted should be maximized subject to runtime constraints.  

\section{Discussion}\label{ch:discussion}
In this section, we discuss the implications of our results, their limitations, and avenues of further investigation. Overall, the results demonstrate the superior efficacy of our methods compared to competing approaches. Our method reliably corrupts the payload while maintaining functional equality, whereas the competing approaches either fail to corrupt the payload or significantly impede network accuracy. Though weight permutation induces a degree of computational overhead, this can be reduced by permuting only set proportions of the model's layers. 

\subsection{The Threat of Neural Network Stegomalware}
Neural network stegomalware is currently sparsely researched despite the significant risks it constitutes. Downloading model checkpoints from unknown sources is relatively common practice both for personal use and in the implementation of machine learning systems, and there is currently no method for verifying whether or not these checkpoints contain a malicious payload. Though the payloads need to be extracted and executed after downloading to take effect, this is not necessarily a significant hurdle in practical settings. For one, there is a general lack of discretion with regards to best practices for evaluating, deploying, and running machine learning models~\cite{large-scale-ml-deployment}. As a result, users may not consider the risks of downloading and executing third-party software. Consider, for instance, an attacker that claims to have implemented an image-generation model that rivals the current state-of-the-art, and that this model can run locally on consumer hardware. In this case, it would not be particularly difficult to distribute both the infected model checkpoint and software that extracts and executes the payload to a large audience. A trivial attack of this nature may for instance be to package the extraction- and execution-code alongside the code for the model architecture in a Pip or Anaconda package. Downloading this package and the checkpoint would complete the attack, and permit the attacker to infect the user with arbitrary malware. This could for instance include ransomware, root-kits, key-loggers, or more tailored forms of malware such as data exfiltration scripts.

Though downloading third-party software in this manner in an of itself constitutes a risk, it has been shown that this is not necessarily required in order to extract and execute the payload due to the many vulnerabilities present in many machine-learning frameworks~\cite{vulnerabilities}. Lui et al.~\cite{stegonet} showed that since neural network checkpoints undergo insecure deserialization, one can feasibly inject custom code to extract and execute the payload in the checkpoint itself. A sophisticated attacker may, as a result, implement an entirely self-contained piece of malware within a neural network checkpoint, which upon loading would extract and execute arbitrary code all while impeding detection. While this vulnerability can be eliminated through the use of Safetensors~\cite{safetensor} or equivalent safe-loading libraries, this is currently far from standard-practice. Moreover, there have also been identified several other vulnerabilities that may permit arbitrary code-execution, therein in the form of uninitialized resource vulnerabilities and misused API calls~\cite{vulnerabilities}.  

Weight permutation effectively mitigates this threat by preventing the extraction of the payload altogether. By shuffling the order of the channel- or column-index of each layer, the payload can no longer be retrieved given the current state-of-the-art methods in neural network steganography. Though the aforementioned vulnerabilities remain, utilizing weight-permutation means that the checkpoint can no longer be used as a medium for stegomalware, and thus that any malicious code may be detected by conventional anti-malware methods such as signature matching.  

\subsection{Limitations}
A significant factor impeding the practical utility of weight permutation is the computational overhead caused by the need for forward-hooks that re-permute the output of each shuffled layer. As shown in \Cref{tab:performance}, these forward hooks add upwards of a 48\% overhead. Though partial permutation can reduce this overhead, as illustrated in \Cref{fig:integrity_sweep}, it may nevertheless be a limiting factor for systems dependent on real-time execution. We plan on investigating optimizations in future work.

As neural network stegomalware is relatively sparsely researched, there is a general lack of understanding with respect to the resilience of neural network stegonography to various common transformations. We have in this work only considered pruning and retraining, as these were the only defenses investigated in the verification of Maleficnet~\cite{maleficnet}. It may, for instance, be the case that Maleficnet is sufficiently mitigated by transfer-learning to certain datasets. Significant architectural modifications such as adding or removing intermediate layers may also impede the extraction process, though this is not necessarily feasible nor realistic for all applications due to its impact on performance. An empirical analysis of the resilience of current approaches to neural network stegomalware to common network operations is as a result warranted.

We also note that weight permutation is not necessarily future-proof. As we will discuss in further detail in the next section, there are several means by which weight permutation can be bypassed. 

\subsection{Future work}
There is significant space for further research with regards to novel embedding methods, neural-network steganalayis, and removal methods.

Though weight permutation effectively mitigates the state-of-the-art in neural network stegomalware, we note that it may be possible to bypass this defense. One possible approach to this is to modify the embedding algorithm to be independent of the order of the weights. This can be achieved in any number of ways, with a simple approach being to perform an initial pass over each layer prior to injection and sort the weight-matrices or convolutional kernels according to the values of their entries. The extraction procedure must then be modified such that the data is sorted back into the proper order, assuming permutation has occurred. It is uncertain, however, whether such a sorting-based approach is resilient to retraining or pruning. 

Another alternative is to develop embedding schemes that are, by design, invariant to permutation. A trivial example of this is to embed the same segments of the payload in every permutable section of a given layer, or equivalently embed it in a channel- or column-wise reduction of the weight tensor. Both of these methods significantly reduce the payload capacity of the network and thus the scope of possible attacks, however. A more sophisticated approach may be to embed the payload in derived representations of the network that are invariant to channel order, but maintain the same dimensionality.

Another potentially impactful extension  may be to increase the redundancy and payload capacity with quantization-aware training~\cite{qat}. This way, the outputs are not significantly affected by large perturbations in the weights, which permits the embedding of more data in the same weight tensor. This increases the scope of possible attacks by virtue of the increased payload capacity, and can reduce the impact of any reductions in capacity that occur as a result of using permutation-invariant embedding schemes. Equivalently, quantization-aware training may serve as a decent defense method. It may for instance be possible to retroactively train a model for quantization and reduce the precision such that any embedded data is corrupted. This is, however, dependent on the efficacy of quantization-aware training methods, and requires more computational resources than can be considered reasonable assuming that the defense is intended to be applied indiscriminately to all instances of checkpoint loading. 

Beyond novel attacks and defenses against stegomalware, there is also a need for further research efforts towards steganalysis methods suitable for neural-networks. For instance, it may be possible to discriminate between infected and clean models by extracting characteristics of the network gradients or the distribution of the least significant weight bits. 

Finally, we highlight that neural network stegomalware is a symptom of the broader problem of lacking security in machine learning systems. This is best exemplified by the numerous vulnerabilities in most machine learning platforms, such as insecure deserialization attacks. Network checkpoints are also distributed without significant attempts at access control on platforms such as, for instance, Pytorch Hub and HuggingFace despite the security concern this represents. We remark that significant research efforts towards improving the safety of both machine learning libraries and platforms is required before neural networks can be safely and effectively utilized in production scenarios. 

\section{Conclusion}
Stegomalware is considered a significant emerging risk to computer security. Neural network stegomalware poses a particularly significant threat to the safety of machine learning systems. State-of-the-art neural network stegomalware is resilient to pruning and retraining, can be extracted and executed with relative ease by leveraging well-known vulnerabilities in machine learning libraries, and can be easily distributed to a wide audience on a variety of popular platforms. Current anti-malware approaches are generally incapable of detecting malicious payloads in neural network checkpoints, and infection merely requires loading a compromised checkpoint into memory. In this paper, we introduced weight permutation, the first effective removal method for neural network stegomalware. By exploiting the permutation symmetry of neural networks and permuting the order of the weight- and channel-columns of each layer, we can effectively corrupt payloads embedded by the state-of-the-art in neural network stegomalware at no cost to network accuracy, outperforming competing methods by a significant margin. Although the practical utility of our method is somewhat impeded by the computational overhead, it can be reduced somewhat at the cost of lower security. We plan on investigating means by which our method can be circumvented in future work, for instance via more sophisticated embedding methods, as well as alternative approaches to payload removal and detection.
\bibliographystyle{IEEEtran}
\bibliography{bibliography.bib}
\end{document}

%% file: glossaries.tex
\newacronym{pd}{WP}{Weight permutation}

%% file: figures/algorithm.tex
\begin{algorithm*}
\caption{Weight Permutation}
\scriptsize
\begin{algorithmic}

\State \textbf{Input:} Neural network $model$
\State \textbf{Method:} $permute\_model\_and\_register\_hooks(model)$
\For{each $conv2d\_module$ in $model$}
    \State $\pi \gets$ random permutation of $conv2d\_module.weight.data$ channel dimension
    \State $conv2d\_module.weight.data \gets \pi(conv2d\_module.weight.data)$ 
    \State $conv2d\_module.bias.data \gets \pi(conv2d\_module.bias.data)$(if applicable)
    \State$conv2d\_module.register\_forward\_hook(\pi^{-1})$
\EndFor
\For{each $linear\_module$ in $model$}
    \State $\pi \gets$ random permutation of $linear\_module.weight.data$ column dimension
    \State $linear\_module.weight.data \gets \pi(linear\_module.weight.data)$
    \State $linear\_module.bias.data \gets \pi(linear\_module.bias.data)$
    \State $linear\_module.register\_forward\_hook(\pi^{-1})$ 
\EndFor
\end{algorithmic}
\label{alg:wp}
\end{algorithm*}

%% file: figures/integrity.tex
\begin{table*}[t]
\centering
\caption{Model Performance under Various Defenses. \checkmark denotes configurations where the payload was corrupted.}
\begin{tabular}{l l l l l l}
\toprule
\textbf{Defense} & \textbf{resnet18} & \textbf{resnet34} & \textbf{resnet50} & \textbf{resnet101} & \textbf{resnet152} \\
\midrule
Weight Shuffling & 1.00 \checkmark & 1.00 \checkmark & 1.00 \checkmark & 1.00 \checkmark & 1.00 \checkmark \\
\midrule
Pruning=0        & 1.00           & 1.00            & 1.00            & 1.00           & 1.00           \\
\midrule
Pruning=0.05     & 1.00           & 1.00            & 1.00            & 1.00           & 1.00           \\
\midrule
Pruning=0.1      & 1.00           & 1.00            & 1.00            & 1.00           & 1.00 \\
\midrule
Pruning=0.25     & 1.00           & 1.00            & 1.00            & 1.00           & 1.00  \\
\midrule
Pruning=0.5      & 1.00           & 1.00            & 0.99            & 0.97            & 0.99           \\
\midrule
Pruning=0.75     & 0.93           & 0.93            & 0.76 \checkmark & 0.79 \checkmark & 0.79 \checkmark \\
\midrule
Pruning=0.9      & 0.43           & 0.14 \checkmark & 0.14 \checkmark & 0.13 \checkmark & 0.13 \checkmark \\
\midrule
Pruning=0.95     & 0.22           & 0.13 \checkmark & 0.13 \checkmark & 0.13 \checkmark & 0.13 \checkmark \\
\midrule
Retraining=1     & 1.00           & 0.99            & 0.97            & 0.97           & 0.98           \\
\midrule
Retraining=5     & 1.00           & 0.99            & 0.98            & 0.98           & 0.97           \\
\midrule
Retraining=25    & 1.00           & 1.00            & 1.00            & 1.00           & 1.00           \\
\midrule
Retraining=50    & 1.00           & 0.99            & 0.99 \checkmark & 0.99           & 0.99           \\
\bottomrule
\end{tabular}
\label{tab:integrity}
\end{table*}

%% file: figures/performance.tex
\begin{table}[t]
\footnotesize
    \centering
        \caption{Permutation defence computational performance. The execution overhead is measured as a percentage of the original execution time, and the loading overhead as the execution time (ms) taken to attach forward hooks and permute the weights in memory.}
    \begin{tabular}{lll}
        \toprule
        Model & Execution Overhead (\%) & Loading overhead (ms)\\
        \midrule
        ResNet18 & 26 & 70 \\
        ResNet34 & 35 & 124 \\
        ResNet50 & 41 & 146 \\
        ResNet101 & 45 & 259 \\
        ResNet132 & 48 & 356 \\
        \bottomrule
    \end{tabular}

    \label{tab:performance}
\end{table}